\documentclass[english,amsmath,amssymb,twocolumn,showpacs]{revtex4}
\usepackage[T1]{fontenc}
\usepackage[latin1]{inputenc}
\usepackage[T1]{fontenc}
\usepackage[latin1]{inputenc}
\usepackage{babel}
\usepackage{graphics}
\date{\today}
\begin{document}
\title{An interplay between nonlocality and quantum violation of path-spin noncontextuality}
\author{Dipankar Home\footnote{dhome@bosemain.boseinst.ac.in}$^1$,
and Alok Kumar Pan\footnote{apan@bosemain.boseinst.ac.in}$^1$}

\affiliation{$^1$ CAPSS, Department of Physics, Bose Institute, Sector-V, Salt Lake, Calcutta
700091, India}
\begin{abstract}
In terms of a suitable variant of the EPR-Bohm example, we argue that the quantum mechanically predicted and experimentally verified violation  of a Bell-type path-spin  noncontextual realist inequality for an `intraparticle' path-spin entanglement involving single neutrons can be used to infer a form of nonlocality, distinct from Bell-type nonlocality, that is required for any relevant hidden variable model to be compatible with the quantum mechanical treatment of an EPR-Bohm-type `interparticle' entanglement. 
\end{abstract}
\maketitle
\section{Introduction}
Spurred on by Bell's seminal work\cite{bell64} based on the EPR-Bohm example\cite{epr, bohmbook}, the study of an incompatibility between quantum mechanics(hereafter QM) and local realist models pertaining to the results of measurements on  the spatially separated particles in entangled states has, for long, become a vibrant research enterprise. Of late, the work of Leggett\cite{leggett} and the subsequent experimental studies\cite{gro,gisin} seeking to rule out a class of nonlocal realist models have stimulated a renewed interest in gaining deeper insights into the relationship between QM and the notion of locality/nonlocality in conjunction with realism\cite{lepu}. On the other hand, for single particles, there has been a considerable body of work on the quantum mechanical violation of noncontextual realism, implying the property which is referred to as `contextuality', viz. that an individual outcome of measuring a dynamical variable, predetermined in terms of a hidden variable model, is dependent upon the measurement (previous or simultaneous) of any other commuting (comeasurable) observable\cite{gleason, bell, kochen, mermin,peres, penrose, roy, cabello, hasegawa1,hasegawa2,home}. 

A connection between the above two strands of investigations is  provided in this paper. For this, we use a suitable example in order to show that the quantum mechanical violation of noncontextual realism for the mutually commuting `path' and spin degrees of freedom of single spin-1/2 particles  enables the inference of a curious form of nonlocality for an EPR-Bohm-type interparticle entangled state of the spatially separated spin-1/2 particles(considered specifically to be neutrons in this paper), a type of nonlocality that is distinct from the Bell-type nonlocality and its various variants. Before going into detailed explanations as given later, the key ingredients of our argument are broadly sketched as follows.

Let us consider the EPR-Bohm-type entangled pairs of neutrons in spin singlets where neutrons in one of the two wings(say, 2) are passed through a particular setup of the Mach-Zehnder-type (Fig.1), while  neutrons in the other wing(say, 1) are subjected to the measurement of, say, either the $z$ or the $x$ component of spin(the measurement settings in the wing 1 for $\widehat\sigma^{1}_{z}$ and $\widehat\sigma^{1}_{x}$ are designated by A and B respectively). In this paper our attention is focused on what happens to  neutrons in the wing 2 corresponding to the two possible settings A and B in the wing 1. 

First, for the setting A, when neutrons in the wing 2 are passed through a specific Mach-Zehnder-type setup, and are separated into two subensembles corresponding respectively to the measurement outcomes $\pm 1$ in the wing 1, the `path' and the spin states of  neutrons belonging to \emph{each} of these subensembles get \emph{entangled}(`intraparticle' entanglement). On the other hand, for the setting B,  after similar such operations, the `path' and the spin states of the neutrons belonging to \emph{each} of the separated subensembles remain \emph{unentangled} while passing through the Mach-Zehnder-type setup in the wing 2. 

Next, note that the separation of neutrons in the wing 2 into two subensembles is conceived essentially on the basis of information obtained about the measurement outcomes in the wing 1. Hence, there is no faster than light signaling involved here. But, what makes this example interesting is the following \emph{contrast} between the cases corresponding to the two settings A and B in the wing 1. 

For the case B, since neutrons belonging to each of the separated subensembles in the wing 2 after passing through the Mach-Zehnder-type setup are in the \emph{path-spin product states}, the contextuality property \emph{cannot} be \emph{discerned} in a Bell-type statistical way for their `path' and the spin degrees of freedom. On the other hand, for the case A, as explained later, because of the path-spin `intraparticle' entanglement generated for neutrons belonging to each of the separated subensembles in the wing 2 while passing through the Mach-Zehnder-type setup, quantum mechanics requires the statistical violation of a  Bell-type path-spin noncontextual realist inequality(NRI)\cite{home} for each of these subensembles, thereby enabling the inference of \emph{path-spin contextuality} for any individual neutron in the wing 2. (The violation of path-spin NRI has been empirically verified by Hasegawa \emph{et al.}\cite{hasegawa1,hasegawa2} using single neutrons). 

Consequently, in this example, in order to be compatible with quantum mechanics, a form of nonlocality is required at the level of individual measured values which are predetermined within any relevant hidden variable model - the meaning of this form of nonlocality, put precisely, is the following: Depending upon the measurement setting in the wing 1, for the case A, for neutrons in the wing 2, path-spin contextuality can be inferred from the statistical measurements, while, for the case B, no such contextuality can be inferred for  neutrons in the wing 2. 

At this stage, before proceeding further, we may remark that the treatment given in this paper, though couched in terms of neutrons, is equally applicable for photons with the appropriate polarizing and analyzing devices. In what follows, the specifics of the required setup and the nuances of our argument are spelled out. We begin with the discussion of the operational ingredients that enable our example to provide a hitherto unexplored slant to the analysis of the EPR-Bohm-type setup. In conclusion, we comment on the significance of such a demonstration of nonlocality.
 
\section{A variant of the EPR-Bohm example}
Let us begin with a source emitting EPR-Bohm-type entangled pairs of neutrons in spin singlets  given by
\begin{equation}
|\Psi\rangle=\frac{1}{\sqrt{2}}\left(|\uparrow_{z}\rangle_{1}|\downarrow_{z}\rangle_{2} -|\downarrow _{z}\rangle_{1}|\uparrow_{z}\rangle_{2}\right)
\end{equation}
where neutrons 1 and 2 are spatially separated into two wings 1 and 2 respectively. In the wing 1, one has the option of making the measurement of either $\widehat{\sigma}_{z}$ or $\widehat{\sigma}_{x}$ (the setting A or B respectively). It  is with respect to these two measurement settings A and B that our subsequent discussions are throughout concentrated on neutrons in the wing 2, while we study the evolution of their spin and path states in passing through a Mach-Zehnder-type setup. Note that, for the measurement settings corresponding to the cases A and B, the mixed states of neutrons 2 can respectively be written as 
\begin{eqnarray}
|\Psi^{A}\rangle_{i}=\frac{1}{\sqrt{2}}\left(|\uparrow_{z}\rangle_{2} \oplus |\downarrow_{z}\rangle_{2}\right)\\
|\Psi^{B}\rangle_{i}=\frac{1}{\sqrt{2}}\left(|\uparrow_{x}\rangle_{2} \oplus |\downarrow_{x}\rangle_{2}\right)
\end{eqnarray}
where $\oplus$ denotes the mixture of two spin states as opposed to their coherent superposition. Since the \emph{same} reduced density operator for the neutrons 2 corresponds to each of the  mixed states $|\Psi^{A}\rangle_{i}$ and $|\Psi^{B}\rangle_{i}$, the statistical properties of any spin variable are the \emph{same} for both these mixed states. The subscript `$i$' is used in Eqs. (2) and (3) to denote the fact that each of these mixed states serves as the initial state that is subjected to subsequent manipulations using a suitable Mach-Zehnder-type setup.

We consider that for each of the measurement settings A and B in the wing 1, the corresponding neutrons 2 are  incident on a 50:50 beam-splitter(BS1) of the Mach-Zehnder-type setup in the wing 2. Any such neutron 2, after passing through BS1,  can then emerge along either the transmitted or the reflected channel corresponding to the state designated by $\left|\psi_{1}\right\rangle$ or $\left|\psi_{2}\right\rangle$ respectively. Subsequently, the neutrons 2 corresponding to $\left|\psi_{1}\right\rangle$ and $\left|\psi_{2}\right\rangle$ are recombined at a second beam splitter(BS2) whose reflection and transmission probabilities are $|\gamma|^{2}$ and $|\delta|^{2}$ respectively. 

Here note that the mutually orthogonal `path' or `channel' states, designated as $\left|\psi_{1}\right\rangle$ and $\left|\psi_{2}\right\rangle$, are  eigenstates of the projections operators $P(\psi_{1})$ and $P(\psi_{2})$ respectively. These projection operators  can be regarded as corresponding to the observables which pertain to the determination of \textit{`which channel'} a particle is found to be in. For example, the results of such  a measurement for the transmitted(reflected) channel with binary alternatives are given by the eigenvalues of $P(\psi_{1})$($P(\psi_{2})$); the eigenvalue $+1(0)$ corresponds to a neutron being found(not found)in the channel represented by $\left|\psi_{1}\right\rangle$($\left|\psi_{2}\right\rangle$). 

Next, a crucial ingredient of our setup is that, after measurement in the wing 1, if the neutrons 2 emerging along one of the two channels, say, represented by $|\psi_{1}\rangle$, pass through a  spin-flipper(SF) that contains a uniform magnetic field along, say,  the $+\widehat x$-axis. Thus, if the spin state of neutron 2 is polarised along $\pm \widehat {z}-axis$, SF \emph{flips} the spin state $\left|\uparrow_{z}\right\rangle_{2}$( $\left|\downarrow_{z}\right\rangle_{2}$) into $\left|\downarrow_{z}\right\rangle_{2}$($\left|\uparrow_{z}\right\rangle_{2}$). On the other hand, if the spin state of neutron 2 is polarised along $\pm \widehat {x}-axis$, the orientation of this spin remains \emph{unaffected} while passing through the SF. 

Subsequently, neutrons 2 in the channels $\left|\psi_{1}\right\rangle$ and $\left|\psi_{2}\right\rangle$ are reflected by the mirrors M1 and M2 respectively - such reflections do not lead to any net relative phase shift between $\left|\psi_{1}\right\rangle$ and $\left|\psi_{2}\right\rangle$. Finally, these neutrons corresponding to $\left|\psi_{1}\right\rangle$ and $\left|\psi_{2}\right\rangle$ are recombined at a second beam splitter(BS2). 

Here we observe that for our subsequent discussions, given the orientation of the magnetic field within SF along the $+\widehat x$-axis, if one considers the spin variables for the two settings of the measurements in the wing 1, one of these  is required to be chosen $\widehat\sigma_{x}$, while the other observable can be \emph{any other} spin variable, which we have taken to be $\widehat\sigma_{z}$ in this specific example.

Now, to proceed with our argument, note that for the measurement setting A in the wing 1 resulting in neutrons 2 with spins polarised along $\pm\widehat{z}$-axis incident on BS1, after the operations in-between BS1 and BS2, the states of these neutrons incident on BS2 are respectively given by
\begin{eqnarray}\nonumber
|\uparrow_{z}\rangle_{2}\rightarrow |\phi_{+}\rangle=\frac{i}{\sqrt{2}}(|\psi_{1}\rangle |\downarrow_{z}\rangle_{2} + |\psi_{2}\rangle|\uparrow_{z}\rangle_{2})\\
|\downarrow_{z}\rangle_{2}\rightarrow |\phi_{-}\rangle =\frac{i}{\sqrt{2}}(|\psi_{1}\rangle |\uparrow_{z}\rangle_{2} + |\psi_{2}\rangle|\downarrow_{z}\rangle_{2})
\end{eqnarray}
where note that, for any given lossless beam splitter, arguments using the unitarity condition show that for the particles incident on a beam splitter, the phase shift between the transmitted and the reflected states of the particles is essentially $\pi/2$\cite{zeilinger}. 

Then, corresponding to the mixed state $|\Psi^{A}\rangle_{i}$ given by Eq.(2), the states of neutrons 2 incident on BS2, represented by $\left|\Psi^{A}\right\rangle_{BS1+SF}$, can be written as 
\begin{eqnarray}
\left|\Psi^{A}\right\rangle_{BS1+SF}=|\phi_{+}\rangle \oplus |\phi_{-}\rangle
\end{eqnarray}
where the constituent subensembles $|\phi_{+}\rangle$ and $|\phi_{-}\rangle$ correspond respectively to the  outcomes $\pm1$ of the measurement of $\widehat \sigma_{z}$ on neutrons in the wing 1. A key point to be stressed is that Eq.(5) denotes a mixture of path-spin \emph{entangled states} where the entanglement arises essentially from the flipping of the spin state $|\uparrow_{z}\rangle_{2}$ or $|\downarrow_{z}\rangle_{2}$ due to the SF placed along one of the channels($|\Psi_{1}\rangle$), in-between the beam splitters BS1 and BS2. 

On the other hand, for the measurement setting B in the wing 1 resulting in neutrons 2 with spins polarised along $\pm\widehat{x}$-axis incident on BS1, the orientations of their spins remain \emph{unchanged} while passing through the SF that contains magnetic field along the $\widehat{x}$-axis. Thus, corresponding to the state $|\Psi^{B}\rangle_{i}$ given by Eq.(3), neutrons 2 incident on BS2  are described by $\left|\Psi^{B}\right\rangle_{BS1+SF}$, given by
\begin{eqnarray}
\left|\Psi^{B}\right\rangle_{BS1+SF}= |\chi_{+}\rangle\oplus |\chi_{-}\rangle
\end{eqnarray}
where the constituent subensembles are given by
\begin{eqnarray}
|\chi_{+}\rangle=\frac{i}{2}\left\{\left|\psi_{1}\right\rangle +  \left|\psi_{2}\right\rangle\right\}\left|\uparrow_{x}\right\rangle_{2}\\
|\chi_{-}\rangle=-\frac{i}{2}\left\{\left|\psi_{1}\right\rangle -  \left|\psi_{2}\right\rangle\right\}\left|\downarrow_{x}\right\rangle_{2}
\end{eqnarray}
which correspond respectively to the  outcomes $\pm1$ of the measurement of $\widehat \sigma_{x}$ on neutrons in the wing 1. In contrast to the state $\left|\Psi^{A}\right\rangle_{BS1+SF}$, the  state $\left|\Psi^{B}\right\rangle_{BS1+SF}$ is a  mixture of path-spin \emph{product states}.
\vskip -1.2cm
\begin{figure}[h]
{\rotatebox{0}{\resizebox{10.0cm}{8.0cm}{\includegraphics{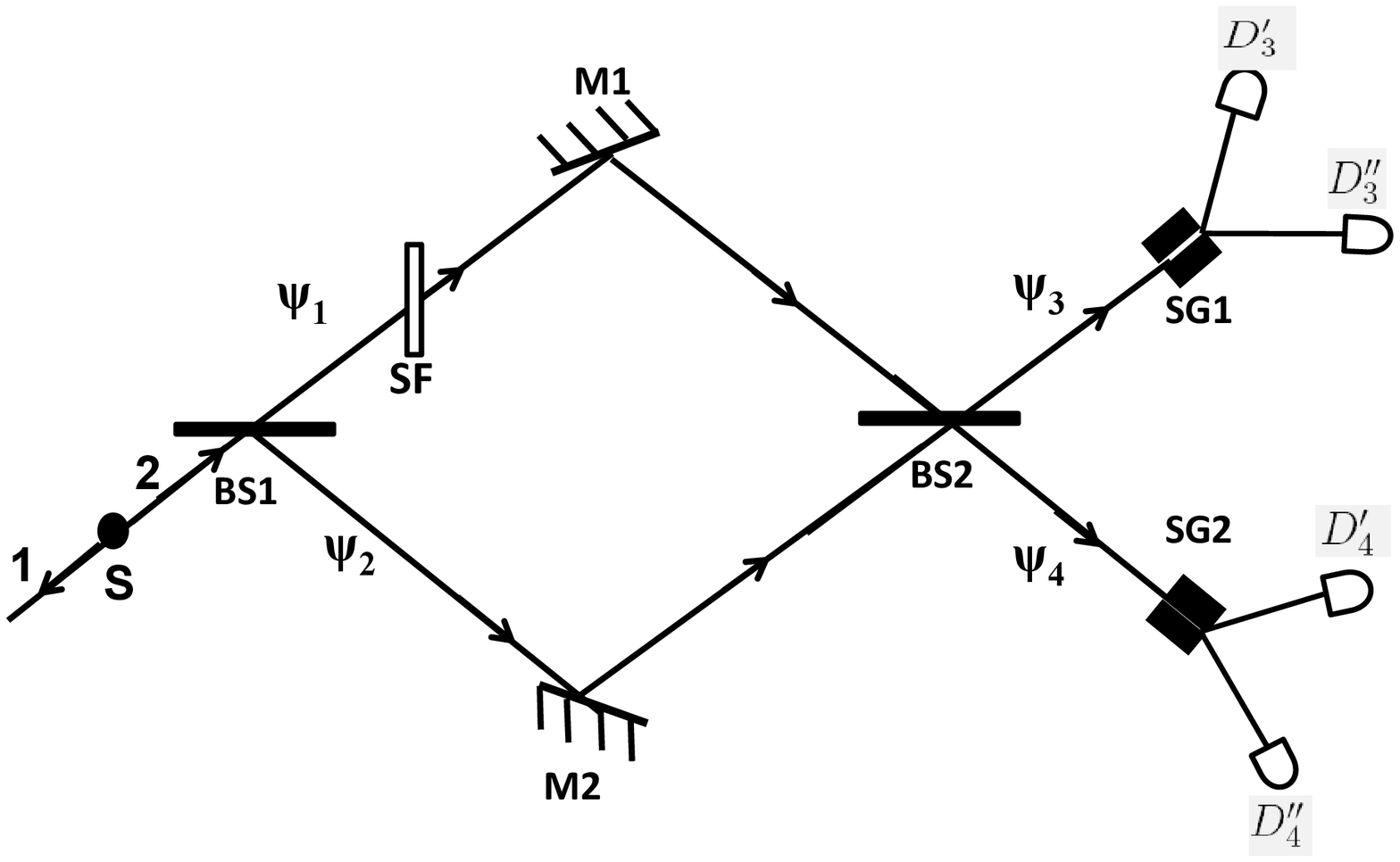}}}}
\vskip -2.2cm
\caption{\footnotesize Neutrons 2 in one of the two wings of the EPR-Bohm correlated pairs emitted by the source(S) enter a Mach-Zehnder-type setup through a beam splitter(BS1) and  pass through the channels corresponding to  $|\psi_{1}\rangle$ and $|\psi_{2}\rangle$. A spin-flipper(SF) is placed along one of the channels $|\psi_{1}\rangle$. Subsequently, appropriate measurements of  the `path' and spin variables are made for these neutrons, as explained in the text.}
\end{figure}
\vskip -0.0cm
Having thus set the stage on the basis of the above setup, we can now develop our argument for showing nonlocality for interparticle entanglement as a consequence of the path-spin contextuality pertaining to `intraparticle' entanglement. For this, it is first important to make the following point. After the corresponding measurements A or B in the wing 1, the process of passing neutrons 2 through the arrangement of BS1+SF serves the purpose of appropriately \emph{preparing} the path-spin states  $|\Psi^{A}\rangle_{BS1+SF}$ or  $|\Psi^{B}\rangle_{BS1+SF}$ respectively on which, subsequently, we  consider the measurements required for  formulating our argument. For this, the beam splitter BS2 of the Mach-Zehnder setup is considered to be part of the arrangement that makes measurements on the neutrons represented by $|\Psi^{A}\rangle_{BS1+SF}$ or  $|\Psi^{B}\rangle_{BS1+SF}$,as explained in the following section.

\section{The relevant measurements and the argument for nonlocality}

We begin by noting that the states of neutrons emerging from BS2, denoted by $\left|\psi_{3}\right\rangle$ and $\left|\psi_{4}\right\rangle$, are unitarily related to the  states $\left|\psi_{1}\right\rangle$ and $\left|\psi_{2}\right\rangle$ by the following relations

\begin{eqnarray}
\label{psi34}
\left|\psi_{3}\right\rangle = -i\gamma \left|\psi_{1}\right\rangle+ \delta  \left|\psi_{2}\right\rangle\\
\nonumber
\left|\psi_{4}\right\rangle = \delta \left|\psi_{1}\right\rangle -i\gamma \left|\psi_{2}\right\rangle
\end{eqnarray}
where $\gamma$ and $\delta$ satisfy  $\gamma^{2}+\delta^{2}=1$.

Eqs.(\ref{psi34} show that, for a given linear combination of  $\left|\psi_{1}\right\rangle$ and $\left|\psi_{2}\right\rangle$, using  different values of  $\gamma$($\delta$), one can generate at the output of BS2 various linear combinations of $\left|\psi_{1}\right\rangle$ and $\left|\psi_{2}\right\rangle$ that correspond to different probability amplitudes of finding neutrons in the channels corresponding to $\left|\psi_{3}\right\rangle$ and $\left|\psi_{4}\right\rangle$. This, in turn, implies that the different values of $\gamma$($\delta$) can be regarded as corresponding to different choices of what may be called the `path' observables $\widehat A_{i}$. Such observables, in terms of actual measurements, correspond to different relative counts registered by the detectors placed along the channels represented by $\left|\psi_{3}\right\rangle$ and $\left|\psi_{4}\right\rangle$. Thus, the beam splitter BS2 plays a key role in this measuring arrangement.

Formally, one can write $\widehat A_{i}=P(\psi_{3})-P(\psi_{4})$ where the eigenvalues $\pm 1$ of $\widehat A_i$ pertain to the detection of a particle in a channel corresponding to either $\left|\psi_{3}\right\rangle$ or  $\left|\psi_{4}\right\rangle$ respectively. Here an important point is that there is an \emph{isomorphism}\cite{takasaki}  between the algebra of the observables $\widehat A_{i}$ and the algebra of $2\times 2$ complex matrices spanned as a linear space by the Pauli matrices $\sigma_{x}$, $\sigma_{y}$, $\sigma_{z}$, and the identity $I$ matrix. This feature can be described as follows.

Taking the representation, for instance, 

$|\psi_{1}\rangle  \rightarrow \left(\begin{array}{c} 
    1\\    0   \end{array}\right)$ ; 
    \hskip 1cm 
    $|\psi_{2}\rangle \rightarrow \left(\begin{array}{c}   0\\  1    
     \end{array}\right)$
     
     and, using the relations given by Eq.(11), it follows that
\begin{equation}
A_{i}=P(\psi_{3})-P(\psi_{4})= \left(\begin{array}{cl} 
    \gamma^{2}-\delta^{2}& \ -i 2\gamma\delta \\  i 2\gamma\delta  & \ \delta^{2}- \gamma^{2} \end{array}\right) 
\end{equation} 
which can be rewritten as the following linear combination of the Pauli matrices 
\begin{equation}
A_{i}=2\gamma\delta \sigma_{y}+ (\gamma^{2}-\delta^{2}) \sigma_{z}= \vec{\sigma}.\vec{a}_{i}
\end{equation}
where $\vec{a}_{i}= 2\gamma\delta \widehat{j}+ (\gamma^{2}-\delta^{2}) \widehat{k}$.

Eq. (11) shows that there is a correspondence between any given `path' observable and a particular component of the Pauli spin vector. As the beam splitter parameter $\gamma (\delta)$ is varied, the `path' observable corresponds to a different component of the Pauli spin vector. It is in this sense that such `path' observables can be regarded as the `\emph{pseudo}-spin' observables that \emph{commute} with the spin observables $\widehat \sigma_{i}$. 

Next, along with the `path' observables $\widehat A_i$, we consider the measurement of the spin variables, say, $\widehat\sigma_{z}$ and $\widehat\sigma_{x}$ using the two suitably oriented Stern-Gerlach(SG1 and SG2) devices(Fig.1)placed along the two output channels $|\psi_{3}\rangle$ and $|\psi_{4}\rangle$ respectively. The detectors associated with SG1 are $D_{3}^{\prime}$ and $D_{3}^{\prime \prime}$, while those with SG2 are $D_{4}^{\prime}$ and $D_{4}^{\prime \prime}$. Registered counts at these respective detectors are denoted by $N^{\prime}_{3}$, $N^{\prime \prime}_{3}$, $N^{\prime}_{4}$ and $N^{\prime \prime}_{4}$. 

Let us concentrate on the joint measurements of the four commuting pairs $\widehat A_{1} \widehat \sigma_{z}$, $\widehat A_{1} \widehat \sigma_{x}$, $\widehat A_{2} \widehat \sigma_{z}$ and $\widehat A_{2} \widehat \sigma_{x}$. The outcome of measuring each of $\widehat A_{1}$, $\widehat A_{2}$, $\widehat \sigma_{z}$ and $\widehat \sigma_{x}$ is $\pm 1$. The respective individual outcomes are denoted by $v(\widehat A_{1})$, $v(\widehat A_{2})$, $v(\widehat\sigma_{z})$ and $v(\widehat\sigma_{x})$. Now, if measurements are considered on a collection of particles that are taken to be prepared in a state which, for a given wave function, is more `completely specified' by a common set of `hidden variables', then, provided `noncontextuality' is assumed,  the following equality holds good for the individual outcomes determined by such `hidden variables': 
\begin{equation}
\label{beq}
v(\widehat A_{1}) v(\widehat\sigma_{z}) + v(\widehat A_{1})v(\widehat\sigma_{x}) +  v(\widehat A_{2})v(\widehat\sigma_{z}) -  v(\widehat A_{2})v(\widehat\sigma_{x}) = \pm 2
\end{equation}
Here it needs stressing that, for the validity of Eq.(\ref{beq}), it is crucial that both the occurrences of, say, $v(\widehat A_{1})$ in Eq.(\ref{beq}) have the \emph{same} value. This means that an individual outcome of measuring the `path' observable $\widehat A_1$ that is predetermined in terms of `hidden variables' is taken to be independent of which spin variable is measured along with it - this is the input of `noncontextuality'. This applies equally for $v(\widehat A_{2})$,  $v(\widehat\sigma_{x})$ and $v(\widehat\sigma_{z})$. 

Next, taking the ensemble averages, it follows from Eq.(\ref{beq}) that
\begin{equation}
\label{bieq}
|\langle \widehat A_{1}\widehat \sigma_{z}\rangle + \langle \widehat A_{1}\widehat \sigma_{x}\rangle + \langle \widehat A_{2}\widehat \sigma_{z}\rangle - \langle \widehat A_{2}\widehat \sigma_{x}\rangle| \leq  2
\end{equation}
Thus, Eq. (\ref{bieq}) can be viewed as an empirically verifiable particular consequence of the noncontextual realist models\cite{home}. A Bell-type inequality  of the form given by Eq.(\ref{bieq}) is what we refer to as noncontextual realist inequality(NRI) which, as applied to our example, is contingent upon the notion that there is no interdependence between the individual outcomes of the spin and the `path' measurements that are predetermined by `hidden variables'.

The violation of NRI, in agreement with the relevant quantum mechanical predictions, has been experimentally demonstrated by Hasegawa \emph{et al.}\cite{hasegawa1} using  single neutrons by preparing an appopriate path-spin entangled state. This shows an empirical violation of noncontextual realism in the sense that if one varies the parameter(namely, $\gamma(\delta)$) that characterizes the context of the `path' or `which channel' measurement for the translational degrees of freedom of a particle, it does affect the outcome of an individual spin measurement. On the other hand, for the path-spin product states, such an effect of `contextuality' does \emph{not} occur since NRI is always satisfied by quantum mechanics . 

Now, we come to the crux of our argument concerning the signature of nonlocality present in the specific variant of the EPR-Bohm example considered in this paper. Here the pivotal point is that a Bell-type path-spin NRI is \emph{violated} separately for each of the two subensembles of neutrons 2 represented by the path-spin entangled states $|\phi_{+}\rangle$ and $|\phi_{-}\rangle$(Eq.(3) and (4) respectively) which are selected corresponding  respectively to the outcomes $\pm 1$ for the measurement of $\widehat\sigma_{z}$ (or, for the measurement of any spin component \emph{other} than $\widehat\sigma_{x}$) on neutrons in the wing 1. On the other hand, for the measurement of $\widehat\sigma _{x}$ on neutrons in the wing 1, any individual neutron in the wing 2 belongs to either of the two \emph{path-spin product states}  $|\chi_{+}\rangle$ or $|\chi_{-}\rangle$(Eqs.(7) and (8) respectively); hence, \emph{no} violation of a Bell-type path-spin NRI given by inequality (13)can be exhibited in this case. The import of this feature concerning nonlocality is as follows. 

Corresponding to the measurement of $\widehat\sigma_{z}$ on  neutrons in the wing 1, one can  regard the quantum mechanical violation of path-spin NRI for each of the two subensembles in the wing 2 to be signifying the property of path-spin contextuality. This means that, in this case, in \emph{contrast} to the case of measuring $\widehat\sigma_{x}$ on neutrons 1, quantum mechanics implies an empirically discernible and statistically manifested interdependence between the individual outcomes of the `path' and the spin measurements which are predetermined in terms of the `hidden variables' for \emph{any} individual neutron 2 - it is in this sense that here a nonlocal effect is required at the level of `hidden variables' in order to be compatible with quantum mechanics. 

Therefore, the upshot of this argument is that, by varying the measurement setting in the wing 1, and by appropriately selecting neutrons in the wing 2 according to the outcomes of the measurements on neutrons 1, one can infer a form of nonlocality on the basis of joint measurements pertaining to the `path' and the spin degrees of freedom of neutrons 2 that test a Bell-type path-spin NRI for each of the two output subensembles. 

Crucial to the above argument is the point that for the measurment setting B ($\widehat\sigma_{x}$ in our example) for the neutrons in the wing 1, sample selection for the neutrons in the wing 2 \textit{cannot} be done in a way that would demonstrate path-spin contextuality through the violation of NRI given by the inequality (\ref{bieq}). In order to understand this point clearly, it needs to be recalled that in the specific interferometric setup used for neutrons in the wing 2, the enclosed magnetic field is oriented along a specific direction, viz. $+\widehat x$-axis within the spin-flipper. Consequently, corresponding to the measurement of $\widehat{\sigma}_{x}$ for the neutrons in the wing 1, since the neutrons in the wing 2 are spin-polarized along either $+\widehat{x}$ or $-\widehat{x}$ axis, there is \emph{no} possibility of the path-spin entangled state being prepared for any individual neutron in the wing 2 that passes through the interferometric setup. Hence, in this case, no matter whatever way one makes the sample selection for the neutrons 2 corresponding to the measurement of $\sigma_{x}$ (the setting B) in the wing 1, the property of path-spin contextuality \textit{cannot} be demonstrated through the violation of NRI given by the inequality (\ref{bieq}) as the `path' and the spin variables remain unentangled for any neutron 2.

In sum, given the experimental setup used in our paper, the path-spin contextuality of neutrons in the wing 2 is statistically manifested (allowing it to be empirically discerned) through the inequality (13) \textit{only} when the measurement A (corresponding to the measurement of any spin component \emph{other than} $\widehat{\sigma}_{x}$) is performed on neutrons in the wing 1, {\it not} when the measurement B (corresponding to the measurement of $\widehat{\sigma}_{x}$) takes place in the wing 1. This is precisely the operational meaning of \emph{nonlocality} that has been argued in this paper.

\section{Concluding remarks} 
We may stress that the type of nonlocality that is shown in this paper is qualitatively \emph{distinct} from the forms of nonlocality that have been demonstrated using various variants(for a useful recent review, see, for example, Genovese\cite{genovese}) of Bell's celebrated theorem. The usual demonstrations of nonlocality for the EPR-Bohm-type entangled states are couched in terms of the correlation properties of the dynamical variables  measured in the two spatially separated wings, where the measurements involved can be spacelike separated. In contrast, our  argument is based on  the measurement of the path-spin correlation properties of neutrons in any one of the two wings, these neutrons being chosen  as belonging to either of the two separate subensembles that are appropriately selected according to the measurement outcomes obtained for the varying measurement settings in the other wing. 

In other words, while the EPR-Bohm-type demonstration of nonlocality violating outcome independence at the level of hidden variables is based on joint measurements of spins in the two spatially separated wings, in our example, by sharing information about outcomes of the spin measurement in one of the two wings, nonlocality  in the form of parameter dependence can be inferred from the joint path-spin measurement on the relevant subsensembes in the other wing.

The argument given in this paper, therefore, leads to the inference that if the constraint of path-spin contextuality is applied to the hidden variable(realist) models  of single spin-1/2 particles, a form of nonlocality is necessarily implied for the hidden variable models pertaining to the EPR-Bohm-type spin entangled states of the pairs of spatially separated spin-1/2 particles. Further studies are called for to analyze closely the nature of this nonlocality, and its connection with Bell-type nonlocality. It may also be worthwhile to explore  whether a variant of the reasoning adopted here can be extended for gaining useful insights into possible constraints restricting the nonlocal realist models in the light of the recent studies\cite{gisin, gro, lepu} based on Leggett's work\cite{leggett}. This is currently being studied.

\section*{Acknowledgements} DH acknowledges the project funding from DST, Govt. of India. DH also thanks the Center for Science and Consciousness, Kolkata for support. AKP acknowledges the Research Associateship of Bose Institute, Kolkata. 

\end{document}